# A Methodology for Discovering how to Adaptively Personalize to Users using Experimental Comparisons

Joseph Jay Williams[1]     Neil Heffernan
[1]HarvardX, Harvard University     Worcester Polytechnic Institute
joseph_jay_williams@harvard.edu     nth@wpi.edu

**Abstract.** We explain and provide examples of a formalism that supports the methodology of discovering how to adapt and personalize technology by combining randomized experiments with variables associated with user models. We characterize a formal relationship between the use of technology to conduct A/B experiments and use of technology for adaptive personalization. The *MOOClet* Formalism [11] captures the equivalence between experimentation and personalization in its conceptualization of modular components of a technology. This motivates a unified software design pattern that enables technology components that can be compared in an experiment to also be adapted based on contextual data, or personalized based on user characteristics. With the aid of a concrete use case, we illustrate the potential of the MOOClet formalism for a methodology that uses randomized experiments of alternative micro-designs to discover how to adapt technology based on user characteristics, and then dynamically implements these personalized improvements in real time.

**Keywords:** experimentation, online education, email, adaptive personalization, methodology

## 1   Discovering how to Adaptively Personalize Technology

A major challenge in adaptive technologies is knowing *how* to adapt – what rules or functions are effective for delivering different versions of content or interactions based on information that varies across users. Often such knowledge for a single technology is obtained from extensive domain experience or years of research into identify domain and student models [1, 5]. Each hour of instructional time provided by the first intelligent tutoring systems could require from 50 to 200 hours of time from PhDs to prepare [8]. Moreover, ill-defined or complex tasks always pose further problems in reliably identifying patterns that can be used for adaptation [5], placing great demands for increasingly detailed models of users [4].

We consider how the MOOClet formalism [11] guides the design and use of technology to conduct experiments [6, 10] in order to dynamically discover how to adapt technologies to different subgroups of users. Specifically, we consider how to unify technology for conducting randomized A/B experimental comparisons with adaptive technologies that deliver different experiences to different users. While there are many examples of experimentation being used in developing and evaluating adaptive web technologies [1], there may be missed opportunities to exploit its value more broadly and deeply.

There has been a rapid increase in the use of experimentation in websites, under the label of A/B or split testing [6]. But the deceptive simplicity of an A/B experiment may obscure that current uses are quite basic in scope and sophistication, relative to the understanding of experimental methodology understanding by behavioral and computational scientists working in laboratories and building systems over the past decades [3]. The rest of the paper considers the relatively untapped opportunity in using experiments to both discover and dynamically implement ways to adaptively personalize technology.

## 2 Unifying Experimentation and Adaptation of modular components: The MOOClet Formalism

The central proposal is to: (1) conduct randomized experimental comparisons of different versions of modular technology components; (2) evaluate the quantifiable benefits of one component over another, *and* to identify whether some components are more or less beneficial for different subgroups of users; (3) dynamically change the policy for  This goes beyond the more frequent use of experiments to determine if one version produces larger impact on a behavior, to a deeper analysis of whether one version is more/less effective than another, based on variation in user characteristics – *different* subgroups of users.

Statistically, this is captured by testing whether the experimental variation of which version is delivered *interacts* with a user characteristic variable, such that the effect of the experimental variable (the difference between the target user behavior based on which version is delivered) *varies* based on the value of a variable representing a user characteristic. That is, for different subgroups of users (defined by different values of the user characteristic) there is a different relative benefit of the different versions. This is a widely used way of analyzing experiments and clearly not novel in itself.

What is novel in the proposed methodology is a systematic and generalizable method for implementing technology to increase the ease and scope of such experimentation, and for turning the discovery of such interactions between an experimental variable and user characteristic directly into rules for adapting technologies to users.

The core insight of this method comes from identifying the formal equivalence between experimentation and adaptive personalization that we characterize below, which further motivates the definition of the MOOClet Formalism. The formalized concept of a MOOClet provides a guide in conceptualizing the design and interpretation of experiments to discover how to adapt technology, as well as a software design pattern. When modular components of technology are implemented using a process that satisfies the formal criteria for such components to be MOOClets, these components support the dynamic mapping of results from experiments into rules for adapting technology.

On the face of it, using technology to conduct a randomized A/B experiment a priori appears to be a very different concept/tool from adapting or personalizing technology to different users. However, there is a way to characterize a formal equivalence between experimentation and personalization for appropriately modular technology components, where both involve delivering alternative versions of a technology component based on the value of variables, differing only in how the values of the variables are set across users.

Consider three alternative versions of an email (or online lesson or exercise), (v1, v2, or vn), one of which is presented to a student based on which of three values of a variable is associated with the student. Figure 1 provides a schematic.

Conducting an experiment comparing the three alternative versions can be formalized as sampling the value of the variable *Expt. 1 Condition* from a random variable that can take on the values A, B, or N. Then, the delivery of versions of a technology component is part of a randomized experiment: version v1 is delivered if Expt. 1 Condition is A, v2 if it is B, or vn if it is N.

Adaptive personalization of the components can be formalized as assigning the value of the variable based on information associated with a user, such as whether the variable *UserCharacteristic1* has the value U1, U2, or UN. Then, the delivery of versions constitutes adaptive personalization: v1 is delivered if UserCharacteristic1 is U1, v2 if it is U2, and vN if it is UN.

In the case of adaptive personalization there is likely substantial existing knowledge about which component to present to which profile or person. On the other end of the continuum, in randomized experiments, it is often unknown which component is effective on average, much less what is effective for particular users.

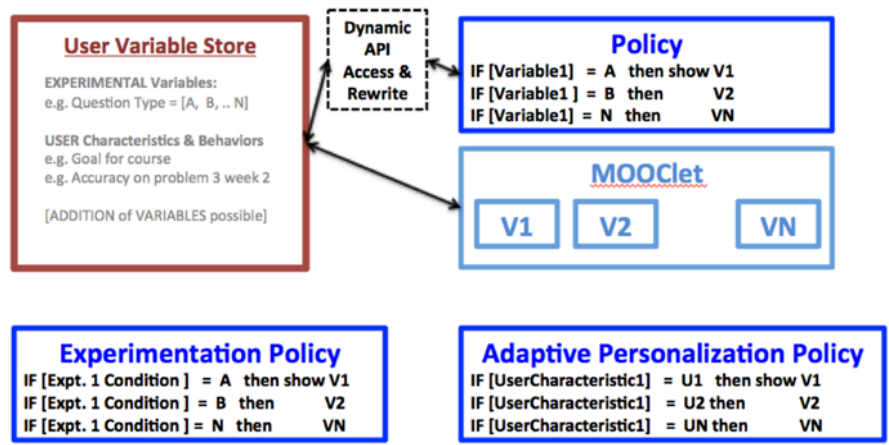

**Figure 1:** Schematic of the MOOClet Formalism. Two examples of using a MOOClet Policy that illustrates how its implementation represents the formal relationship between experimentation and adaptive personalization.

**Formal Definition of a MOOClet.** A modular sub-component of a technology (e.g., exercise, lesson) is implemented a a MOOClet **if and only if**:
1. Can be modified to produce multiple versions – labeled **MOOClet-Versions.**

2. When a MOOClet is accessed by a learner, it can select one of the multiple MOOClet-Versions to be presented by accessing and conditioning on the values of a set of variables associated with a user. This set of variables is termed a **User Variable Store**. The set of rules for which MOOClet-Version is delivered conditional on variables in the User Variable Store is labeled a **MOOClet-Policy**.

4. The User Variable Store is not fixed, but can be dynamically updated by changing the values of existing variables or by adding new variables.

5. The rules in the MOOClet-Policy is not fixed, but can be dynamically updated to editing or adding the variables that are used to condition on and the logical rules that determine which version is presented based on these variables.

## 3 Usage Scenario: Experimentation and algorithms to discover effective adaptive personalization rules for emails based on age and activity

One Usage Scenario is now presented for how the MOOClet Formalism can improve technology by guiding experimentation to discover how to adaptively personalize technology. We do not report the specific experimental data about behavioral outcomes, as these are in preparation with collaborators for publication in an archival format. The "results" in this section are therefore details about how the methodology was successfully implemented in a specific real-world context.

In a HarvardX Massive Open Online Course (MOOC), there was a plan to send out emails asking non-active participants about why they had stopped engaging in the course. The goal for sending these emails was to collect information (not to reengage participants) and so the quantitative outcome to be maximized by applying the MOOClet Framework was to increase response rate to these emails.

Instead of using the default mailer used by instructors and researchers at HarvardX, the email message was implemented according to the criteria and design pattern of a MOOClet. That is, different versions of an email message could be delivered using a MOOClet-Policy of IF-THEN rules based on variables in a User Variable Store. Both the variables in the User Variable Store and the IF-THEN rules used could be dynamically rewritten or added to, essential in this usage scenario.

The initial creation of different MOOClet-Versions, initialization of experimental variables in the User Variable Store, and definition of the MOOClet-Policy logic was to conduct an experiment, which consisted of three independent experimental manipulations that compared different versions of the email. Each experimental variable had three different conditions. For example, the subject line of emails was varied, so that there were three different subject lines presented. The other (independent) experimental variables were three different versions of the introductory message, and three different prompts for people to click a hyperlink within the email (to answer questions about why they were no longer participating). In all, there were 27 different versions of the email, based on a combination of the 3 different subject lines x 3 different introductory messages x 3 different prompts to respond to a hyperlink.

In addition, information about participants' age and number of days active in the course so far (e.g. 0, 1, or 2 or more days) was added to the User Variable Store. The age variable was transformed to a new discretized age variable within the User Variable Store, binning it into five roughly equally sized categories (e.g. 18-22, 23-26).

Therefore, implementing the emails as MOOClets allowed for embedding an experiment within the original plan to simply send out emails to elicit participant responses. In addition to the 3 x 3 x 3 experimental design, there was data about the associated user characteristics of age and number of days active, and the infrastructure for changing how emails were delivered as a function of these variables, which could be done at any time. In particular, it could be done based on analyzing the results of the initial experiment, data for which was passed into the User Variable Store in real time.

A round of emails was sent out to approximately 4000 participants, and data about whether and what they had responded was passed back to the User Variable Store for real time analysis. The target variable to optimize was the proportion of people responding to an email. Statistical tests found that this varied as a function of all experimental variables (subject line, intro message, prompt to follow hyperlink), and that some of these variables also interacted with user characteristics of age and number of days active.

For sending out emails to approximately a further 1500 participants, 500 participants were sent out using the same method was used in running the experiment – 33.3% of participants in every one of the three conditions, randomly assigned independently for each of the three experimental variables.

However, to allow adaptive personalization while continuing to experiment, a different algorithm was used to update the proportion of people randomly assigned to each condition. Originally, these weights or proportions were 33.3% in every experimental condition (there were 3 conditions in each variable). However, in the subsequent emailing, while random assignment was still used to determine which participant was assigned to each condition, an algorithm was used to adjust the weights or proportions of participants in a given condition that were used in sampling condition for each new participant.

This algorithm weighted the assignment of a condition in proportion to how high response rate had been to that condition, relative to the others. Moreover, to do adaptive personalization, those weights were computed independently for each value of the user characteristic variables (i.e., age group, and number of days active).

Response rate was increased by more than half in this adaptive personalization condition that used data about interactions of the conditions from the 3 x 3 x 3 experiment with age group and number of days active. In addition to this practical benefit, the statistical power to detect the effect of the experimental variables was still increased by continuing to assign people using this partially random method. Simple regressions statistics that control for the fact that assignment was partially random, but also influenced (weighted) by user characteristics, were used to confirm these gains to statistical power (a discussion of these statistics is beyond the scope of this particular report).

More discussion of relevant algorithms is contained in [12], which also explains how the MOOClet Formalism provides an abstraction for reinforcement learning agents: The different versions of a MOOClet are the actions a reinforcement learning agent can take, the reward function is the target variable in the User Variable Store identified for being optimized by varying the MOOClet Version, and the (optional) state space is a chosen subset of the variables representing user characteristics in the User Variable Store. Any technology component implemented using the MOOClet formalism can be modified in real-time via relatively simply APIs by applying algorithms for multi-armed bandits, contextual bandits, and (Partially Observed) Markov Decision Processes [2, 7, 9, 12].

## Acknowledgements

We acknowledge support through NSF Cyberinfrastructure Award 1440753, SI2-SSE: Adding Research Accounts to the ASSISTments Platform: Helping Researchers do Randomized Controlled Studies with Thousands of Students.